
\input harvmac

\Title{\vbox{\hbox{HUTP--94/A005,}}
       \vbox{\hbox{IASSNS-HEP-94/16}}      }
{\vbox{\centerline{Mirror Transform and String Theory}}}
\vskip .2in

\centerline{C. Vafa}
\vskip .2in \centerline{Lyman Laboratory of Physics, Harvard
University}
\centerline{Cambridge, MA 02138, USA}
\centerline{and}
\centerline{School of Natural Sciences, Institute for Advanced Study}
\centerline{Olden Lane, Princeton, NJ 08540, USA}
\vskip .3in
Some aspects of Mirror symmetry are reviewed, with an emphasis
on more recent results extending mirror transform to higher
genus Riemann surfaces and its relation to the Kodaira-Spencer
theory of gravity\foot{Talk given in the {\it Geometry and Topology
Conference}, April 93, Harvard, in honor of Raoul Bott.}.

\Date{3/94}
One of the most beautiful aspects of string theory is that
strings moving on one manifold may behave identically with
strings moving on a different manifold.  Any pair of manifolds
which behave in this way are called mirror pairs.  In this paper
I will review some aspects of this phenomenon
(see \ref\mirb{Contributions to {\it Essays on Mirror Manifolds},
ed. by S. T. Yau,  International Press, 1992.} for a collection
of articles on this subject).

One of the most successful methods in solving difficult
questions in mathematical physics has been to transform
the problems from their initial setup to a more trivial
setup where everything can be understood in simple terms.
One of the most classical examples of this is Fourier
transform which transforms a differential equation to a polynomial
equation which is easy to solve. A more recent such example
is the twistor transform which maps the classification of
self-dual geoemetries and self-dual Yang-Mills fields in four
dimensions to the question of classification of appropriate
complex manifolds and holomorphic vector bundles in six
(real) dimensions which is more manageable.   The most recent
example of such a transformation is the mirror transform
which is the subject of this paper.
The idea is that one maps certain difficult
questions of interest
in algebraic geometry on one manifold to simple questions
of variation of Hodge structures on a different (mirror) manifold.
As we shall see, however,
 the status of this transformation is not as clearly
understood as the other two
cases we mentioned.

We first need to recall some basic aspects of string theory.
The main ingredient in a string theory is a two dimensional
quantum field theory.  An interesting class of 2d QFTs
are sigma models, which may be viewed as path integrals
over the space of maps $\phi$ from a Riemann surface
$\Sigma ^g$ (of genus $g$) to a manifold $M$:
$$\phi: \qquad \Sigma^g \longrightarrow M$$
The integral over the space of maps
is written abstractly as
$$\int D\phi e^{-S(\phi )}=F_g(M)$$
where $S(\phi)$ is the energy functional $S(\phi)=\int_{\Sigma^g}
|d\phi|^2$
and $F_g(M)$ is called
the genus $g$ partition function of strings propagating on $M$.
In most cases of interest in superstrings one considers
supersymmetric sigma
models which means that in addition to maps from the Riemann surface
to $M$
one considers fermionic fields which take their values in the tangent
bundle to $M$ and one modifies $S$ in an appropriate way to obtain
 a supersymmetric theory.

We say that strings {\it cannot} distinguish between $M$ and $M'$, or
$M \equiv M'$ if and only if\foot{I am giving a definition of
equivalence
which is necessary but has not been proven to be sufficient, though I
believe
it is also sufficient. The strict definition of equivalence will have
to include not just the partition function but all the correlation
functions.}
$$F_g(M)=F_g(M')\lambda^{g-1}\qquad for \quad all \quad g$$
for some constant $\lambda$ (playing the role of string coupling
constant).
The simplest example of mirror symmetry corresponds to choosing
$M$ to be a circle of circumference $L$ and $M'$ to be a circle
of circumference $1/L$.  This case is particularly easy because
$F_g(M)$ can be explicitly computed (this is seldom the case
in more interesting examples). Identifying the map $\phi$ from
$\Sigma^g$
to the circle with the coordinate on the circle, we have
$$\phi \sim \phi +L \qquad for \quad M_1$$
$$\phi \sim \phi +{1\over L} \qquad for \quad M_2$$
The space of maps from the Riemann surface to the circle
decomposes into infinitely many components depending on how the
surface wraps around the circle.  To be concrete, let
$a_i, b_i$ as $i$ runs from 1 to $g$ deonte a canonical
basis for $H_1(\Sigma^g)$, and let $\alpha^i$ and $\beta^i$
be the corresponding harmonic one forms.
Then for each set of integers $(n_i,m_i)$ we get a component
of the map of the surface to $M$ characterized by the condition that
$$d\phi =(n_i \alpha^i +m_i \beta^i) L+d {\tilde \phi} $$
where ${\tilde \phi}$ is a univalued function on $\Sigma$.
Expanding the action $S[\phi]$ for this component we get
$$S[\phi]\rightarrow  L^2\int (n\alpha +m \beta)\wedge *(n\alpha
+m\beta)
+S[{\tilde \phi}]$$
Now we can consider the $L$ dependence of $F_g$.  We will have
to sum over all disconnected components of such maps.  However the
fact that for each component $S$ decomposes in the way described
above
we see that
$$F_g(L)=\big(\sum_{n,m}
exp[-\int (n\alpha +m\beta)\wedge * (n\alpha +m \beta)]\big)
\cdot \int D{\tilde \phi} exp[S[{\tilde \phi}]]$$
Note that the path integral we are left to perform
is independent of $n,m$. Even
 though the path integral over $\tilde \phi$
can be computed easily the important aspect to emphasize
in our case is its dependence on $L$. The only dependence on $L$
comes from
the constant maps ${\tilde \phi}$ where it gives a factor of $L$
(the rest of it just gives us $(det'\Delta/\int \sqrt g )^{-1/2}$
where $\Delta$ is the Laplacian on the Riemann surface).
We thus immediately deduce that
$${F_g(L_1)\over F_g(L_2)}={L_1 \sum_{n,m}
exp[-L_1^2 \int(n\alpha +m\beta)\wedge
*(n\alpha +m\beta)]\over L_2 \sum_{n',m'}
exp[-L_2^2 \int(n'\alpha +m'\beta)\wedge
*(n'\alpha +m'\beta)]}$$
Note that if $L_2=1/L_1$ then this ratio is equal to
$(1/L_1^2)^{g-1}$ by using
Poisson resummation on $(n,m)\rightarrow (n',m')$.
We have thus seen that
$$F_g(L)=F_g(1/L) \lambda^{g-1}$$
where $\lambda =1/L^2$.
This implies that two circles with circumferences which
are inverse to each other are mirror pairs. One can easily extend
this example to the target being a $d$-dimensional torus.  This
mirror
symmetry has actually been used to give a model
of string cosmology
\ref\bvt{R. Brandenberger and C. Vafa, Nucl. Phys. B316 (1989)
391\semi
A. Tseytlin and C. Vafa, Nucl. Phys. B372 (1992) 443}.
  One of the main troubles
of early universe cosmology for point particle
theories is that the universe
would be singular at the time of the big bang.  This gives
rise to many unphysical things such as infinite temperature
and infinitely strong gravitational fields.   In the context
of strings a toroidal universe will not have these difficulties
since it maps a singular universe
(a torus with zero size) to infinite size universe
which is manifestly non-singular.

In the context of superstrings one usually considers target
spaces being a
$d$-complex dimensional
 K\"ahler manifold which admits a Ricci-flat
metric.  Such K\"ahler manifolds are known as Calabi-Yau
manifolds (
K\"ahler manifolds with trivial canonical line bundle which
admit a nowhere vanishing holomorphic $d$-form).
Of particular importance in studying sigma models
on Calabi-Yau is the structure of the moduli space. There are
two types of deformations to consider:  Complex deformations and
K\"ahler deformations.  The complex deformations of Calabi-Yau
relevant for sigma models is the classical one (studied
in \ref\titod{
G. Tian, in {\it Mathematical aspects of String theory, ed. by S.
T. Yau}, World Scientific, Singapore, 1987 \semi
A.N. Todorov, Comm. Math. Phys. 126 (1989) 325.}).
 The dimension
of this moduli space is equal to the dimenison of $H^1(M,T)$ which
for Calabi-Yau manifolds is the hodge number $h^{1,d-1}$.
  The K\"ahler deformations
in the context of sigma models is more complicated
than the classical picture and in particular belong to the
complexified $H^{1,1}(M)$. The real part of it plays the usual
role of a K\"ahler class whereas the complex part plays
the role of introducing a phase in the measure of the sigma model.
The basic idea is that if $b$ denotes the complex part of the
K\"ahler class this means that we modify the path integral
measure $D\phi$ by
$$D\phi \rightarrow D\phi \ {\rm exp}(i\int_{\Sigma}\phi^*(b))$$
where $\phi^*(b)$ is the pullback of
the two form $b$ to the Riemann
surface. The reason it is called the complexified K\"ahler
class is that it can be unified with the real part to
write the action in a way which automatically includes
the above twisting of the path integral measure. Let
${\tilde g}_{i\bar j}=g_{i\bar j}+i b_{i\bar j}$.
Let $\phi^i$ denote the map $\phi$ in component form.
Then the action can be written as
$$S=\int {\tilde g}_{i\bar j} \partial \phi^i \bar \partial
\phi^{\overline j}+{\tilde g}^*_{i\overline j}\bar \partial \phi^i
\partial
\phi^{\overline j}$$
It often is convenient to formally think of $\tilde g$ and
${\tilde g}^*$ as independent parameters and take the
${\tilde g}^*\rightarrow \infty$.  In this case the
second piece of the action blows up
for generic maps and the path
integral will be finite only for holomorphic maps
where the second piece vanishes.  In this way
the asymmetric limit of fixing $g$ and sending
${\tilde g}^*\rightarrow \infty$ gives a path-integral
which basically measures how many holomorphic maps
there are from the Riemann surface to the
K\"ahler manifold\foot{Strictly speaking one needs to consider
the topologically twisted sigma model to get this
\ref\wit{E. Witten, Comm. Math. Phys. 117 (1988) 353;
 Nucl. Phys. B340
(1990) 281.}.}.

Let us consider a simple example of one dimensional
Calabi-Yau manifolds.  This is nothing but the elliptic
curve.  The moduli space of this manifold as far
as the complex deformation is concerned is
parametrized by one complex parameter $\tau$
on the upper half plane.  One
represents the torus in the standard fashion as a parallelogram
with sides $1$ and $\tau$ imbedded in the complex plane.
To get the moduli space we have to recall that the $\tau$'s differing
by
 the $PSL(2,{\bf Z})$ action represent the same torus.  As far
as the Ricci-flat K\"ahler metrics are concerned, the area $A$
of the flat torus is the only parameter.  However as discussed
above this is complexified
to $A-iB$.  Let us define the complex parameter $\rho$
parametrizing complexified K\"ahler deformations by
$$\rho =iA+B$$
Note that since $B$ affects the path-integral only by
multiplication by phases it is a periodic variable as far
as the moduli space of the theory is concerned.  So we have
to identify
$$B\sim B+1$$
which means that
$$\rho \sim \rho+1$$
There is in fact a larger symmetry to mod out in order
to get the moduli space: Let us consider a rectangular
torus of size $(R_1,R_2)$ on each side.  Setting $B=0$
this gives $\rho =iR_1R_2$.  From
the discussion about the duality of the circle, it
should be clear that if we consider a rectangular
torus with sides $(1/R_1,1/R_2)$  we get an
equivalent theory: This gives $\rho =i/(R_1R_2)$.
So we see that in this case we have the
following identification
$$\rho \sim {-1\over \rho}$$
It can be shown that this applies even to more general
configurations of the torus, and so the moduli space
of K\"ahler deformation, is identified with $\rho$ up
to the group generated by these two transformations
which magically enough is again $PSL(2,{\bf Z})$.
We therefore see that $\tau$ and $\rho$ play
very similar roles in this example, even though one
parametrizes the complex structure and the other parametrizes
the K\"ahler structure.  In fact as it turns out to get
the full moduli space there is an extra symmetry to mod
out and that is the $\tau \leftrightarrow \rho$ exchange.
To see this consider the special case of rectangular
torus discussed above.  To begin with we have
$$(R_1,R_2)\rightarrow \tau ={iR_2\over R_1} \qquad \rho =iR_1R_2$$
We can consider an equivalent theory by taking the duality
transformation on the first circle.  This leads to
$$({1\over R_1},R_2)\rightarrow \tau'=iR_2 R_1 \qquad
\rho'={iR_2\over R_1}$$
which means that there is an exchnage symmetry of $\tau$ and $\rho$.
This can be shown to be more generally valid for any torus and
leads to the first example of the mirror phenomenon.
Below we shall see that this is a special case of the general
phenomenon of Calabi-Yau mirrors where the role of complex and
K\"ahler deformations are exchanged.

Even though the above example is very instructive it is a bit
too special and one would like to find more non-trivial examples
of the above phenomenon.  It turns out there is at least a hint:
In distinguishing manifolds we consider topological invariants.
For example the Euler characteristic $\chi$.  We conclude that if
$$\chi(M_1)\not= \chi (M_2)\rightarrow M_1\not=M_2$$
So we can easily disprove the existence of the isomorphism
between $M_1$ and $M_2$ if their Euler characteristic is different.
Similarly for string theory we should try to construct {\it
invariants of }$2d$ {\it QFT} in order to distinguish them.
If two QFT's have different such invariants then they
cannot be mirror pairs.  These invariants are to be constructed
out of objects which canonically make sense in the
algebraic formulation of QFT's without any recourse to a target
manifold.
For supersymmetric QFT's which
are the ones which arise in considering superstrings propagating
on K\"ahler manifolds, there is a basic inavariant which is
Witten's index
$$I=Tr(-1)^F$$
where $(-1)^F$ is a mod 2 gradation of the Hilbert space
of the QFT (by using the fermion number) and the trace
is over the full Hilbert space.  $(-1)^F$ is characterized
by the fact that it squares to one and it anticommutes
with the operator which implements supersymmetry $Q$.
However this makes $(-1)^F$ well defined only up to an overall
sign change
$$(-1)^F\rightarrow -(-1)^F$$
We thus see that only $|I|$ is an invariant for the QFT.
If two QFT's arising from sigma models on two manifolds
have different $|I|$'s we can deduce that they cannot be
mirror pairs.  In the case of a supersymmetric sigma model
on a manifold $M$ it turns
out that \ref\witeul{E. Witten, Nucl. Phys. B202 (1982) 253.}
$$|I| =|\chi (M)|$$
We therefore deduce that if $|\chi(M_1)|\not= |\chi(M_2)|$
then $M_1$ and $M_2$ cannot be mirror pairs.  This however
leaves open the door for two manifolds being topologically
distinct and having $\chi(M_1)=-\chi(M_2)$ but which
are nevertheless mirror pairs; at least the index $|I|$
does not distinguish them.  Note that this condition
 for finding mirror pairs fits very well with
the example of torus considered above because
as we saw torus was its own mirror and on the other hand the
Euler characteristic of torus is zero.

We could try to ask if there are analogs of more refined
invariants such as the betti numbers, or for the
case of the Calabi-Yau manifolds, which is of most
interest to us, the hodge number $h^{p,q}$, which can
distinguish different sigma models.  It turns out that
$h^{p,q}$ can be defined as the dimension of a canonical
subspace of the Hilbert space of the QFT (the ground
state subsector with a particular $U(1)\times U(1)$ charge).
But there is still an ambiguity.  Let $d$ be the complex
dimension of the Calabi-Yau manifold.  Define
$${\hat p}=-{d\over 2}+p$$
Then there is a canonical $U(1)\times U(1)$ gradation
of the Hilbert space of the corresponding QFT.
Let ${\cal H}^{\hat p, \hat q}$ denote the subspace
of the Hilbert space consisting of ground states with
$U(1)\times U(1)$ gradation given by $\hat p ,\hat q$.  Then
$$h^{p,q}={\rm dim}{\cal H}^{\hat p,\hat q }$$
However again it turns out that there is a $Z_2$ ambiguity
in canonicaly assigning the $U(1)\times U(1)$ gradation which
is obtained by flipping the relative sign of the two $U(1)$'s:
$(\hat p,\hat q)\rightarrow (-\hat p ,\hat q)$.
This in particular means that the dimension of
a given ${\cal H}^{\hat p, \hat q}$
could in principle correspond either to $h^{p,q}$ or
$h^{d-p,q}$.  Just as before the existence of this
ambiguity leaves the door open for having mirror pairs
$(M_1,M_2)$ for which in particular
$$h^{p,q}(M_1)=h^{d-p,q}(M_2)$$
$$\chi (M_1)=(-1)^d \chi (M_2)$$
It was conjectured based on the existence of simple
toroidal examples and the lack of any method to
resolve the $Z_2$ ambiguity in computing $h^{p,q}$
that this kind of mirror pairs always
exist \ref\mir{W. Lerche, C. Vafa and N. P.
Warner, Nucl.Phys. B324 (1989) 427
\semi L. Dixon in Proc. of the 1987 {ICTP Summer Workshop in
High Energy Physics and Cosmology}, Trieste}.

Note that for a Calabi-Yau manifold the dimension of
moduli space of complex deformations is given by $h^{d-1,1}$
(this is proven in \titod ) and the K\"ahler
deformations by $h^{1,1}$.  Therefore for mirror pairs
the two moduli spaces interchange their roles: The complex
deformations for one correspond to K\"ahler deformations for
the other and vice-versa.

The mirror conjecture as stated cannot be possibly true
because there exist rigid Calabi-Yau manifolds
with no complex deformations. So their
mirrors will have no K\"ahler deformations, i.e., it will not
even be a K\"ahler manifold.  However there is a generalization
of the conjecture (involving supermanifolds) which can take
care of such cases \ref\seva{S. Sethi and C. Vafa, in preparation.}.

An important element in relating
($N=2$) supersymmetric 2d QFT's (and in particular
conformal field theories) with geometry of Calabi-Yau was
Gepner's construction \ref\gep{D. Gepner, Phys. Lett. B199 (1987)
380.}.
These were CFT's which were constructed
by taking the tensor product of simplest representations of
($N=2$) Virasoro algebra (minimal models)
which had an unexpected similarity with what
one would expect for CFT's coming from sigma models on certain
Calabi-Yau manifolds.
Subsequently it was discovered that the Landau-Ginzburg
description of ($N=2$) minimal models
\ref\vwm{C. Vafa and N. Warner,
Phys. Lett. B218 (1989) 51.\semi E. Martinec, Phys. Lett. B217 (1989)
431.}\
associates an isolated singularity
(taken as the superpotential of Landau-Ginzburg model) to each
minimal model.
The simplest ones correspond to monomials $x^n$.  Thus
the Gepner's construction was reinterpreted in this context
as taking combinations of these monomials and constructing
quasi-homogeneous polynomials to be identified with the defining
equation of the associated Calabi-Yau manifold in weighted projective
space \ref\gvwm{B. Greene,
C. Vafa and N. Warner, Nucl. Phys. B324 (1989) 371.}\ (see also
the recent work \ref\witcy{E. Witten, Nucl. Phys. B403 (1993) 159.}).

{}From this point on one could use amazing properties
of conformal field theories to construct mirror pairs.  In particular
if one considers the minimal model given by $x^n$, the theory
has a $Z_n$ symmetry given by multiplying $x$ by an $n$-th root
of unity.  It turns out that if we divide out this $Z_n$ symmetry
we end up with the original theory again!  This is somewhat
unexpected as usually one expects that if we divide out a theory
by a symmetry we should get a theory with fewer degrees of freedom.
This is not the case for conformal theories because dividing by
a symmetry gets rid of some states which are not invariant but
at the same time introduces new states coming from loops which
are closed only up to the group action.  So the total degrees of
freedom remain constant.  That we end up with the same theory
in this case can be seen by realizing that this model
has an associated circle corresponding to phases of
$x$ and modding out by $Z_n$ changes its radius from $R$ to
$1/R$, and thus the previous argument applies to show that
we end up with the original theory.

First non-trivial examples of Calabi-Yau mirrors were
constructed in \ref\grpl{B. R. Greene and M. R.
Plesser, Nucl. Phys. B338 (1990) 15}\ using this symmetry.
In particular it was shown there that if we consider
the quintic given by $\sum_{i=1}^5 x_i^5=0$ in $CP^4$ and
modding out by the maximal subgroup of $Z_5$ phase multiplications
of each monomial consistent with preserving the holomorphic 3-form
($Z_5\times Z_5 \times Z_5$) one gets the same theory back but
now interpreted as a sigma model on a quotient of the quintic
three-fold.   There was also
a large class of examples suggested by computation of the
Euler class of Calabi-Yau threefolds (as hypersurfaces in
weighted projective space) in which for almost all manifolds searched
of Euler class $\chi$ there was one with Euler class $-\chi$
\ref\graph{P. Candelas, M. Lynker, R. Schimmrigk, Nucl.
Phys. B341 (1990) 383.}.
There has been further work giving a conjectured construction
for mirror pairs for a very wide class of examples
\ref\batbor{V.V. Batyrev, {\it Dual polyhedra and mirror
symmetry for Calabi-Yau hypersurfaces in toric varieties},
preprint Universit\"at-GHS-Essen, 1992 (to apppear in J. Alg.
Geom.)\semi
V.V. Batyrev and D. van Straten, {\it Generalized Hypergeometric
Functions and Rational Curves on Calabi-Yau Complete Intersections
on Toric Varieties}, preprint 1993\semi
L.A. Borisov, {\it Towards the Mirror Symmetry for Calabi-Yau
Complete
Intersection in Gorenstein Toric Fano Varieties}, preprint 1993
(alg-geom 10000193).}.

The  existence of mirror pairs as discussed up to now seems
like a pure curiosity as far as string theory is concerned.
It simply points out that topological invariants
are not necessarily good stringy invariants.  However what
makes the existence of mirror pairs mathematically interesting
is that certain computations of QFT correlation functions in
one manifold which have one mathematical interpretation for
one manifold have a completely different interpretation
for the other.  Therefore the existence of mirror symmetry implies
in particular that two different mathematical computations
on two distinct manifolds are unexpectedly equal.  Moreover
it turns out that on the one side the computation is simple
and on the other side no one knows how to perform it!  The
mirror pairs thus serve in this context as a way to
transform a difficult computation to an easy one.

The difficult computation is what is known as the computation of
quantum
cohomology ring \wit ,
  which is a deformation
of the classical cohomology ring. The deformation parameters
are in one to one correpsondence with K\"ahler deformations
of the manifold.  In the limit we take the size of the manifold
to infinity the quantum cohomolgy ring reduces to the classical
one.  The simplest way to desribe it is to note that classical
cohomology ring can be described by counting the intersection
{\it pionts} of dual cycles representing the cohomology classes.
The quantum cohomlogy ring is simply a {\it blurred} version
of classical intersection theory where the intersection {\it point}
is
replaced by a rational curve in the manifold.  We say three
cycles intersect if they all intersect a given rational curve.
We weigh each such intersection by the number of points they
intersect the rational curve and
in addition by ${\rm exp}(-\int k)$
where $k$ is the (complexified) K\"ahler class and it is integrated
over
the rational curve in question.
Note that if we let $k\rightarrow \infty$ the rational curves
are suppressed in this quantum intersection theory
and so we are left with the degree zero rational
curves which are just points, i.e. we get back the classical
intersection theory.

The hard computation is thus a coholomolgy ring deformation which
is a function of K\"ahler moduli.  The general structure
of mirror pairs discussed above suggests that the mirror
computation should involve rings which are functions
of complex moduli. It turns out that such objects have
already been encountered in the mathematics under the general
subject of studying variations of Hodge structure.
For the particular case of the Calabi-Yau threefold the
mirror of the quantum triple intersection defined above
can be computed as follows:  For each cycle dual to the
K\"ahler class on one manifold there is a direction for the
variation of complex structure on the mirror manifold.  So
fixing three classes in the previous computation will correspond
to fixing three directions for the variation of complex sturcture
on the mirror, which we denote by $(i,j,k)$.
 Let $\omega$ be a non-vanishing holomorphic 3-form
of the Calabi-Yau manifold.  The mirror computation is simply
written as
$$C_{ijk}=\int \omega \wedge \partial_i \partial_j \partial_k \omega
$$
This is easy to compute and so we have managed to transform
a difficult computation of quantum triple intersection
on one manifold to classical questions of variations of Hodge
structure
on the other.  The first explicit example was carried out
in \ref\candet{P. Candelas, Xenia C. de la Ossa,
Paul S. Green and L. Parkes, Nucl. Phys. B359 (1991) 21}\
 for the quintic three fold
and led to the impressive
computation of the number of rational curves of arbitrary
 degree on the quintic.

One can continue this line of correspondence in the context
of arbitrary genus Riemann surface.  On the difficult side
one has to compute the number of holomorphic maps from a
genus $g$ Riemann surface to the Calabi-Yau threefold;
more precisely the natural question is to allow the moduli
of the Riemann surface to vary and ask how many holmorphic
maps there are for some point on the moduli space of
the Riemann surface.  Actually for high enough genus
one expects not to get isolated holomorphic maps
but families of them.  In such cases the natural
thing to compute is the Euler character of a certain
bundle over the moduli space of holomorphic maps
from Riemann surfaces of genus $g$ to Calabi-Yau manifolds
\ref\witt{E. Witten, Nucl. Phys. B371 (1992) 191.}.

This is the hard part of the question and we would like to
discuss what is the mirror computation which
is expected to be simple? This question has recently
been answered \ref\bcov{M. Bershadsky, S. Cecotti, H. Ooguri
and C. Vafa, Harvard University preprint, HUTP-93/A025.}\
 and we shall now summarize
some of the results of this work.

At the level of sphere the easy side of the computation
corresponded to classical computations
which means ordinary integrals which arise even for
point particle theories.  In other words the sphere
computation, which is the tree level for strings,
leads to  classical computations
(ordinary integrals) which in principle could
arise from point particle theories
(the usual overlap of wave functions).  Therefore it is
natural to expect that the higher genus computations
which are quantum corrections for string theory
will be mapped to quantum corrections for the
would be point particle theory.  This expectation is borne
out, and the corresponding point particle theory
which quantizes the variations of Hodge structure
is called the Kodaira-Spencer theory of gravity \bcov .

The basic idea of this theory is to start with the classical
equations of the theory which is the variations of complex
structure of a Calabi-Yau threefold $M$.
According to Kodaira-Spencer theory, the variation of the complex
structure can be encoded by saying how $\bar \partial$ varies:
$$\bar \partial \rightarrow \bar \partial +A \partial$$
where $A$ is a section of $TM \times {\overline T^*}M$, which
I will denote, counting the vectors with negative weight as opposed
to forms by a (-1,1) `vector-form'.  The Kodaira-Spencer equation
which is the consistency condition for
the new $\bar \partial$ for a finite shift of complex structure
 holomorphic tangent bundle is
$$\bar \partial A+ {1\over 2}[A ,A]=0$$
where the commutator takes the commutator of the A as a vector
and wedges the antiholomorphic forms.  To solve this equation
we follow \titod . We use
the notation that when we contract a vector-form with
the holomorphic three form $\Omega$ we put a prime over the
vector-form.  Contracting $A$ with $\Omega$ gives us a (2,1)
form $A'$.  One chooses variations $A$ which in addition respect
the three form, which means we choose $A$ such that
$$\partial A' =0$$
Then the the lemma of \titod\ shows that in such a case
$$[A,A]'=\partial (A \wedge A)'$$
By contracting the KS equation with $\Omega$
(which is nowhere zero) we get the equivalent equation
$$\bar \partial A' +{1\over 2}\partial (A\wedge  A)'=0$$
For a physical realization,
this has to be the classical equations of an action which
is indeed the case.  The action is given by
$$S={1\over 2}\int_M  A' {1\over \partial}\bar \partial A'
+{1\over 6} \int_M A' \wedge (A\wedge A)'$$
This action including the condition $\partial A'=0$ arises
naturally in string field theory \bcov .  The action
as written above is formal because of ${1\over \partial}$ but
it can be made precise sense of \bcov .
Tree level expansions
of this theory naturally gives rise to perturbative
solutions of the KS equation which had been studied
before \titod .
  The fact that the
tree level computations of this theory leads to the
computations which arise in the variations of Hodge structure
is not surprising--after all the tree level computations
are equivalent to classical equations which in this case
is the KS equation.

One should then go on to compute the quantum corrections for the
above
action.  In such cases one typically encounters divergencies
which should be regularized.  The first such case is the one-loop
correction.  For the Kodaira-Spencer action the one-loop correction
turns out to be a certain particular combination of holomorphic
Ray-Singer torsions.  Let $\Delta_{p,q}$ be the Laplacian
acting on $(p,q)$ forms on $M$.  The one-loop correction
turns out to be\foot{The stringy version of this is
$F_1={1\over 2}\int {d^2\tau \over \tau_2} Tr(-1)^F F_LF_R
q^{H_L}{\bar q}^{H_R}$ where the integral is over the moduli
space of torus, $q$ is the modular parameter, $H_{L,R}$ denote
the left- and right- moving Hamiltonians, and $F_{L,R}$ denote
the $U(1)\times U(1)$ left-right fermion number gradiation.
This object was defined in \ref\cv1{S. Cecotti and C. Vafa,
Comm. Math. Phys. 157 (1993) 139.} as a generalization
of Ray-Singer torsion to the loop space of K\"ahler
manifolds.}
$$F_1={1\over 2}
{\rm log} \prod_{p,q} {\rm det}\Delta_{p,q}^{pq(-1)^{p+q}}$$
of course for such an object to make sense one
needs to delete the zero modes and in addition it
 needs to be regularized.  This
kind of determinant has been rigorously studied in
 \ref\bis{J.M. Bismut and D.S. Freed,
Comm. Math. Phys. 106 (1986) 159; 107 (1986) 103\semi
J. -M. Bismut, H. Gillet and C. Soule,
Comm. Math. Phys. 115 (1988) 49, 79, 301
\semi J.M. Bismut and K. K\"ohler, {\it Higher analytic
torsion forms for direct images and anomaly formulas}, Univ. de
Paris-sud, preprint 91-58.}\
and regularized using the zeta
function regularization techniques\foot{To the best of my knowledge
the particular combination of torsions appearing
here which leads to a magically simple anomaly formula
for the Calabi-Yau manifolds
has not been studied in the mathematics literature.}.
In order to compute $F_1$
one generally studies how $F_1$ varies as a function of moduli
of complex structures of Calabi-Yau manifold.  Formally
one argues that $F_1$ is a sum of a holomorphic
and an antiholomorphic function on moduli space.  But as is well
known
there is an anomaly, the Quillen anomaly,
 which falsifies the formal argument.  This anomaly is captured
by studying the curvature
 $\partial \bar \partial F_1$.  This can be computed either
using string techniques \ref\bcovb{M. Bershadsky, S. Cecotti,
H. Ooguri and C. Vafa, Nucl. Phys. B405 (1993) 279.}\ or the more
conventional
techniques \bis\ with the result that\foot{
In the stringy computation the  ${1\over 24}$ in the second term
arises
in computation as a result of the volume of the moduli space of
torus whereas in the index derivation it appears in the expansion
of todd class.}
$$\partial \bar \partial F_1={1\over 2}\partial \bar \partial
[\sum_{p,q}
(-1)^{p+q} q d_{p,q}]+{1\over 2}\int_M Td(T)\sum_p(-1)^pp
\ Ch(\wedge^{n-p} T^*)
\big|_{(1,1)-{\rm part}}$$
$$={1\over 2}\partial \bar \partial [\sum_{p,q}
(-1)^{p+q} q d_{p,q}]-{\chi (M)  \over 24}G$$
where $d_{p,q}$ denotes the determinant of harmonic $(p,q)$
forms (in a holomorphic basis), $Td,Ch$ denote the Todd and
chern classes, $T$ and $T^*$ denote the holomorphic tangent and
cotangent bundles, $\chi (M)$ denotes the Euler characteristic
of $M$, and $G$ is the K\"ahler form on moduli space (which
is the same as $c_1({\cal L})$ where $\cal L$ is
the line bundle over the moduli space whose
section is a holomorphic 3-form).  In evaluating
the index integral  above use has been made of the fact
that Calabi-Yau manifolds have vanishing first chern class.
Integrating the anomaly equation to get
 $F_1$ is now an easy task with the additional boundary condition
of how $F_1$ behaves near the boundaries of moduli space.
It should be noted that $F_1$ defined above makes sense
for CY manifolds of arbitrary dimension.

Let us consider the simplest example of mirror phenomenon, namely
the elliptic curve, and compute $F_1$ in either language. In the
easy side of the computation as we discussed above we are
going to get the analytic torsion which is simply
$$F_1(q)=-{\rm log}{\sqrt \tau_2} \eta(q) \overline{\eta (q)}$$
where $q={\rm exp}(2\pi i \tau )$ specifies the complex structure
of the torus and $\eta$ is the Dedekind $\eta$ function given by
$$\eta =q^{1\over 24}\prod_{n=1}^{\infty}(1-q^n).$$
Of course the easy way to show this
is to study Quillen anomaly as discussed above:
Applying the analysis to this case is rather simple and we end up
with the anomaly equation that
$$\partial_\tau \bar \partial_{\bar \tau} F_1= {-1\over 2(\tau -\bar
\tau)^2}
.$$
Modular invariance and regularity in the interior gives
rise to the $\eta$ function as we integrate the
above anomaly equation.
On the `difficult' mirror to this as we mentioned
we would expect to be counting the number of holomorphic curves
from the torus to the manifold which in this case is again
a torus where
as discussed before
now the role of $\tau$ and $\rho$ are exchanged--in other
words we now think of $\tau$ as parameterizing the complexified
K\"ahler class of the torus.
  Fixing a point on the torus and mapping it
to a particular point on the target torus (dual
to the K\"ahler class) is equivalent to studying
$q\partial F_1/\partial q$.  Note that in the dictionary
for the mirror map $q$ counts the degree of the holomorphic
map to the manifold, so the coefficient
in front of $q^n$ signifies the number of maps of degree $n$.
However the existence of anomaly means that we have
an inevitable mixture of $q$ and $\bar q$.  To simply count
the holomorphic curves we should consider
an asymmetric limit fixing
$g$ but taking $g^*\rightarrow \infty$ which
concentrates the path integral on holomorphic maps
as discussed before.
In this case this means keeping $\tau$ fixed and sending $\bar \tau
\rightarrow \infty$.  We thus have (expanding the logarithm
of the Dedekind $\eta$ funtion)
$$q{\partial F_1\over \partial q}\big|_{\bar \tau \rightarrow
\infty}={-1\over 4\pi i(\tau -\bar \tau)}\big|_{\bar \tau \rightarrow
\infty}+\sum_{n,m=1}^{\infty} nq^{nm}-{1\over 24}=
\sum_{n,m=1}^{\infty} nq^{nm}-{1\over 24}$$

Let us check whether this is what we expect by directly
counting the number of holomorphic maps from torus to torus:
At degree zero, every elliptic curve can be mapped holomorphically
to the torus by the constant map. So at degree zero we have
a continuous family parameterized by the moduli space of tori.
The physics computes (up to sign) roughly
the volume of this space which in this case
is ${1\over 24}$.  The coefficients in front of
$q^{nm}$ should signify the number of holomorphic curves
which cover the target torus $nm$ times.  Let us start with
$n=m=1$. For a fixed target torus, there is clearly a unique
torus for which there exists a holomrphic map, namely the
torus with exactly the same modulus as the target torus and the
map being the identity map.  This explains the coefficient
of $q$ being 1.  We can consider an $nm$-fold covering of this
torus, by going $n$ times covering in one direction and
$m$ times covering in another direction.  By tilting
the parallelogram representing the torus in one direction we
can actually get $n$ inquivalent such tori.  One can show
by a careful check that up to $SL(2,Z)$ transformation there
are no more (in particulr the number of holomorphic
maps of degree $r$ is equal to $\sum_{n\ {\rm divides}\ r}n$)
in agreement with the above formula derived by using mirror
symmertry.

We can also ask how higher quantum corrections can be computed.
As remarked before higher loop corrections correspond
to maps from higher genus Riemann surfaces to the Calabi-Yau
manifold.  It is at this point that threefolds play a very
distinguished
role, in that the quantum corrections vanish for all Calabi-Yau
$n$-folds except for $n=3$.  On the `easy side', which for higher
loops are not so easy, one is computing the quantum corrections
to the Kodaira-Spencer theory.  This will have to be studied
very carefully with an eye on regularization of the theory.  On
the hard side, one is computing the number of holomorphic maps
(or the Euler character of an appropriate bundle on moduli
space of holomorphic maps)
from genus $g$ Riemann surface to the three fold.

So the question is how one does the computation?  The
method followed in \bcov\ makes use of the fact
that the answer is expected to be essentially
a holomorphic function of moduli (or more precisely
a holomorphic section of an appropriate line bundle
on moduli space).  If it were exactly holomorphic to completely
fix it one would only need to know its
behavior near the boundaries of moduli space
which in general would require only a finite data.  However the
holomorphicity is not exactly right at higher genus
just as it was
not exactly right for genus one, where one encountered
Quillen's anomaly. In the genus one case, to get a
well defined answer as mentioned above we have to take
the holomorphic derivative of $F_1$ in the direction
of moduli and the Quillen anomaly is the statement that
this function is not a holomorphic function of moduli.
For higher genus $F_g$ is well defined without having to take
any derivatives, because the higher genus does not have
any isometries, so the anomaly is going to be captured
by a statement involving ${\bar \partial}F_g \not= 0$.
In order to compute this anomaly, one has in principle
two options: Either use the Kodaira-Spencer theory directly
and study its anomaly at higher loops (which is the generalization
of what has been studied in the case of one loop in \bis ), or
study it directly in string theory where $F_g$ is defined
as the integral of a particular measure on moduli space of Riemann
surfaces
of genus $g$.  Since the regularization of Kodaira-Spencer
theory has not been studied at higher loops the first approach
seems difficult at present.  It turns out that the second approach,
i.e. using string theory techniques, is extremely easy.
Even if the regularization of Kodaira-Spencer theory
had been studied it is hard to believe the derivation of the
anomaly would have been as simple as the stringy derivation.
In the context of string theory one finds that
$$\bar \partial_i F_g= \int_{{\cal M}_g} \partial \bar
\partial \rho_{\bar i}$$
where ${\cal M}_g$ is the moduli space of Riemann surfaces
of genus $g$, and $\rho_{\bar i}$ is a well defined
form on moduli space.  Using this structure the computation
reduces to the contributions from the
 boundary of moduli space of Riemann surfaces. Naturally
objects one encounters there would be related to lower
genus computations.  The answer one obtains turns out to be
$$\partial_{\bar i} F_g =
 {1 \over 2} \overline{C}_{\bar{i}\bar{j}\bar{k}}
 g^{j \bar j} g^{k \bar k}
 \left( D_j D_k F_{g-1} + \sum_{r=1}^{g-1}
          D_j F_r \ D_k F_{g-r} \right)$$
The two boundary contributions on the right hand side
come from the two distinct type of degenerations of the
Riemann surface (from the handle degeneration and splitting
of the surface respectively).  Without going to too much
detail let us just note that $F_g$ is a section of ${\cal L}^{2-2g}$
where $\cal L$ is the line bundle of $H^{3,0}(M)$ on the moduli
space, and $D_i$ represent covaraint derivatives
with respect to the natuarl connection
 and $g_{i\bar j}=\int_M \omega_i \wedge {\overline \omega_j}$
 represent the canonical metric on the
$(2,1)$ forms $\omega_i$.
This anomaly equation can be naturally captured in the master
form including all genera at once by considering
$Z={\rm exp}\sum_{g=1}^{\infty}\lambda^{2g-2}F_g$
and we have
$$\big[ \overline{\partial}_{\bar i} - \overline{\partial}_{\bar i}
F_1 -{\lambda^2 \over 2} \overline{C}_{\bar{i}\bar{j}\bar{k}}
 g^{j \bar j}
     g^{k \bar k}    {D}_{j} {D}_k\big]Z=0$$

So the basic strategy to compute $F_g$ at higher genus
is by induction. Suppose
we know $F_g$ up to a given genus.  To get the next one, we
integrate the above anomaly formula
 using the lower genus $F_g$ and the behaviour of $F_g$ near the
boundaries of moduli space.  Using this strategy the
case of the quintic threefold was studied in \bcov\ with the
result indicated in the table below.

\vfill
\eject

\vskip .5in

\centerline{\vbox{\offinterlineskip
\hrule
\halign{&\vrule#&
   \strut\quad\hfil#\quad\cr
height2pt\cr
&Degree\hfil&&$g=0$&&$g=1$&\cr
height2pt&\omit&\cr
\noalign{\hrule}
height2pt&\omit&\cr
&n=0&&5&&50/12&\cr
&n=1&&2875&&0&\cr
&n=2&&609250&&0&\cr
&n=3&&317206375&&609250&\cr
&n=4&&242467530000&&3721431625&\cr
&n=5&&229305888887625&&12129909700200&\cr
&n=6&&248249742118022000&&31147299732677250&\cr
&n=7&&295091050570845659250&&71578406022880761750&\cr
&n=8&&375632160937476603550000&&154990541752957846986500&\cr
&n=9&&503840510416985243645106250&&324064464310279585656399500&\cr
&...&&...&&...&\cr
& large n && $a_0n^{-3}(\log n)^{-2} e^{2 \pi n \alpha}$
&& $a_1n^{-1} e^{2 \pi n\alpha}$ &\cr
height2pt&\omit&\cr
\noalign{\hrule}
height2pt&\omit&\cr
\noalign{\hrule}
&Degree\hfil&&$g=2$&&$g$&\cr
height2pt&\omit&\cr
\noalign{\hrule}
height2pt&\omit&\cr
&n=0&&-5/144&&-$100\cdot [c_{g-1}^3]$&\cr
&n=1&&0&& &\cr
&n=2&&0&& &\cr
&n=3&&0&& &\cr
&n=4&&534750&& &\cr
&n=5&&75478987900&& &\cr
&n=6&&871708139638250&& &\cr
&n=7&&5185462556617269625&& &\cr
&n=8&&90067364252423675345000&& &\cr
&n=9&&325859687147358266010240500&& &\cr
&...&&...&&...&\cr
& large n && $a_2 n (\log n)^2 e^{2 \pi n\alpha}$ &&
$a_g n^{2g-3} (\log n)^{2g-2} e^{2 \pi n \alpha}$ &\cr
height2pt&\omit&\cr}\hrule}}

\centerline{Table 1. \#\ curves of genus $g$ on quintic hypersurface}

\vfill
\eject
In this table we have included also the genus 0 answer which
was computed in \candet\ and the genus 1 result which was computed
using the anomaly in \bcovb .
As far as the numbers which have been checked mathematically, up to
now
the genus 0 answer up to degree $3$ has been confirmed.
For higher genus, the direct computations are even more difficult.
The only non-trivial one which has
been confirmed is by Stromme et. al. \ref\stro{Private
communication.}\
(using a formula of Bott) and it is for the $n=4$
for $g=1$ (the $n=3$ for $g=1$ can be easily argued to be
equal to $n=2$ for $g=0$ which agrees with the physical
computation).  For $g>2$ we have determined only the asymptotic
behavior of the number of holomorphic curves
(to fix all the numbers we have to have a little bit more precise
knowledge about the finite number of coefficients which
dominate the divergence behavior of $F_g$ near the boundaries
of moduli space of the quintic and fixes the holomorphic
ambiguity in integrating the anomaly equation).

Mirror symmetry is a reflection of the rich structure
in the loop space of Calabi-Yau manifolds
in that they give rise to 2 dimensional conformal theories
with $N=2$ supersymmetry.  As is well known
these are only very special examples of $N=2$
superconformal theories, many of which have no target manifold
interpretation.  We have learned from the existence of mirror
symmetry that classical concepts of invariants are to be
modified when going to loop space, and in fact we are
led to ask more generally what are the invariants of the $N=2$
conformal
field theories which arise for CY sigma models.
  Having a good understanding of these
invariants will be a necessary tool in classifying them.
So far only very limited set of invariants of the $N=2$
theories have been constructed.  There could very well be
more refined invariants (by invariants we mean
aspects of the theory which are unaffected by
moving in the moduli space of these theories).  In particular
defining invariants over integers which is natural
for manifolds has not yet been defined for the $N=2$ CFT's.
This is a challenging and important area to develop.

I would like to thank the hospitality of the Institute for Advanced
Study at Princeton where this paper was written.

This research is supported in part by a fellowship from
Packard foundation and NSF grants PHY-89-57162 and
PHY-92-18167.

\listrefs

\end